\documentclass[twocolumn,twoside,showpacs,amsmath,amssymb,byrevtex,altaffilletter,citeautoscript,jasa]{revtex4asa}

\usepackage{graphicx}
\usepackage{dcolumn} 
\usepackage{bm}      
\usepackage{helvet}  

\begin{document}
\title{Resonance modes in a 1D medium with two purely resistive
boundaries: calculation methods, orthogonality and completeness}
\author{Jean Kergomard}
\email{kergomard@lma.cnrs-mrs.fr}
\author{Vincent Debut}
\affiliation{Laboratoire de M\'{e}canique et d'Acoustique, CNRS
UPR 7051, 31 Chemin\\ Joseph Aiguier, 13402 Marseille Cedex 20,
France}
\author{Denis Matignon}
\affiliation{T\'{e}l\'{e}com Paris/D\'{e}partement TSI, CNRS UMR
5141\\ 37-39, rue Dareau 75014 Paris, France}
\defineheaderauthor{Kergomard}
\definerunningtitle{Resonance modes in a 1D medium}
\date{\today }

\begin{abstract}
Studying the problem of wave propagation in media with resistive
boundaries can be made by searching for ``resonance modes'' or
free oscillations regimes. In the present article, a simple case
is investigated, which allows one to enlighten the respective
interest of different, classical methods, some of them being
rather delicate. This case is the 1D propagation in a homogeneous
medium having two purely resistive terminations, the calculation
of the Green function being done without any approximation  using
three methods. The first one is the straightforward use of the
closed-form solution in the frequency domain and the residue
calculus. Then the method of separation of variables (space and
time) leads to a solution depending on the initial conditions. The
question of the orthogonality and completeness of the
complex-valued resonance modes is investigated, leading to the
expression of a particular scalar product. The last method is the
expansion in biorthogonal modes in the frequency domain, the modes
having eigenfrequencies depending on the frequency. Results of the
three methods generalize or/and correct some results already
existing in the literature, and exhibit the particular difficulty
of the treatment of the constant mode.

\end{abstract}
\pacs{43.20Ks, 43.40Cw, 02.30Tb, 02.30Jr} \maketitle

\section{Introduction}

Studying the problem of wave propagation in media with resistive
boundaries
can\ be made by searching for ``resonance modes'' (see Ref.%
\onlinecite{filippi}), or free oscillations regimes. These modes
can be non-orthogonal for the ordinary scalar product, entailing
some difficulties depending on the mathematical treatment, made
either in the time or frequency domain. Two classical methods
exist for such a problem, and can be used either for a scalar,
second order differential equation, or for a system of two
equations of the first order. They have been especially used for
the problem of a 1D medium with one resistive boundary, the other
boundary condition being of Dirichlet type:

i) in the time domain, the use of time and space variable as
separate variables leads directly to the basis of modes, but they
are non-orthogonal for the most common product, and difficulties
occur when searching for the coefficients depending for instance
on initial conditions. Nevertheless, for a particular case,
Oliveto and Santini \cite{oliveto}, and Guyader \cite{guyader}
have solved the problem, and Rideau \cite{rideau}, using a system
of equations of the first order, found a
scalar product making the modes orthogonal (see also Refs. %
\onlinecite{int,cox,DVD}), and gave the proof of completeness.

ii) in the frequency domain, the equations to be solved are
ordinary differential equations with boundary conditions depending
on frequency, but the use of orthogonal decomposition is possible.
This leads to eigenmodes and eigenfrequencies {\it depending on
frequency.} It is the case for the classical theory of room
acoustics (see e.g. Morse and Ingard\cite{morse} ), using
biorthogonality. To return to time domain in order to deduce the
resonance modes is a rather delicate task, especially because of
the calculation of the derivation of eigenfrequencies with respect
to frequency. Biorthogonality has been used also for duct modes
(see e.g. Ref. %
\onlinecite{och})  Another approach has been recently used by
Trautmann and Rabenstein \cite {TRO,TRO2}, using a system of first
order equations (these authors treat the case of two resistive
boundary conditions).

The present article is devoted to the study of the simple 1D case,
when the two boundaries are resistive. One goal is to exhibit how
the different methods articulate. We start by using the fact that
a straightforward solution exists for the wave equation with
source, by applying the residue calculus to the closed-form of the
Fourier domain solution: as discussed by Levine\cite{levine}, this
closed-form solution, avoiding the sum of a series, is
``relatively poorly, if not entirely, unknown to the general
acoustics community''. All calculations can be carried out
analytically without any approximation, exhibiting the properties
of the different methods (however many previous papers restrict
their content to small impedance, or admittance, at one extremity,
using perturbation methods). The case under study corresponds to
one-dimensional propagation in a homogeneous medium bounded by two
other semi-infinite media with different characteristic
impedances, dissipation being therefore due to radiation at
infinity. It is especially interesting because of its physical
significance (it is probably the simplest radiation problem), and
also because it realizes one of the possible transitions between
Neumann and Dirichlet boundary conditions. Notice that in the
context of optics and quantum mechanics, the problem has been
studied including the outside media by Leung
et al\cite{leung0,leung}, the resonance modes being called quasinormal modes.%
\newline
In section \ref{statement}, the equations to be solved are stated,
with some possible physical interpretations. As a first step, the
classical, closed-form solution of the Green function in the
frequency domain is established (section \ref{compact}), with its
inverse Fourier Transform, corresponding to the successive
reflections (section \ref{successive}). The second step is the
residue calculus in order to determine the resonance
modes (section \ref{inverse Fourier}, the basic result being given by Eqs.~(%
\ref{29})). Then results are compared to those of the two
aforementioned methods, i.e.: i) the method of separation of
variables (section \ref {separation}), which gives the result for
given initial conditions (the corresponding results being
Eqs.~(\ref{35}), (\ref{49}) and (\ref{50a})); in this section, the
question of orthogonality and completeness of the modes is
investigated. ii) the method of eigenmodes in the frequency domain
(section \ref{eigenfrequencies}). For the two methods, both second
order scalar equation and first order system of two equations are
used successively, with emphasis on the existence of a constant
mode.

\section{Statement of the problem, physical interpretation\label{statement}}

The Green function $g(x,t\mid x_{0},t_{0})$ for the wave equation
is solution of the following equation:
\begin{equation}
\left[ \partial _{xx}^{2}-c^{-2}\partial _{tt}^{2}\right]
g(x,t)=-\delta (x-x_{0})\delta (t-t_{0})  \label{z1}
\end{equation}

where $x$ and $x_{0}$ are the spatial coordinates of the receiver
and source, respectively (or vice-versa), $t$ and $t_{0}$ the
times of observation and excitation, respectively, $c$ the speed
of sound. $\delta (x) $ is the Dirac function.

For sake of simplicity, $x_{0}$ and $t_{0}$ are considered to be
fixed.
Moreover in the whole paper, the choice of $t_{0}=0$ is made. For negative $%
t,$ the function is zero, as well as its first derivative. The
Green function satisfies the following boundary conditions:

\begin{eqnarray}
c\zeta \partial _{x}g(x,t)=\partial _{t}g(x,t)\text{ \ at
}x=0\text{,} \label{z2} \\ c\zeta _{\ell }\partial
_{x}g(x,t)=-\partial _{t}g(x,t)\text{ \ at }x=\ell \label{z3}
\end{eqnarray}

where $\zeta =Z/\rho c$, $\rho $ is the density of the fluid, and
$Z$ the impedance at $x=0,$ which is assumed to be a real
quantity, independent of the frequency. Similarly, $\zeta _{\ell
}=Z_{\ell }/\rho c$ , where $Z_{\ell }$ is the impedance at
$x=\ell $ ($\ell $ being positive).

\begin{figure}

\ifgalleyfig   
  \includegraphics[width=3.25in]{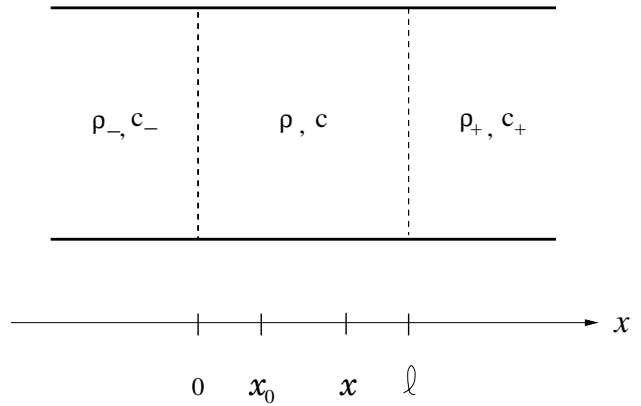}
\else
  \ifoutputfig  
  \includegraphics[width=5.5in]{fig1.eps}
  \fi
\fi

\caption [1D medium bounded with two other media.] {\label{fig1}
\ifgalleyfig {1D medium bounded with two other media.} \fi}
\end{figure}

 An obvious physical interpretation for quantities $\zeta $
and $\zeta _{\ell }$ is the following: consider for $x<0$ and
$x>\ell $ \ (see figure 1) two media with characteristic
impedances $\rho _{-}c_{-}$ and $\rho _{+}c_{+}$, respectively. If
the media are non dissipative, impedances are real, and can be
larger or smaller than the impedance of the bounded medium, $\rho
c$. Moreover, they are positive, because they correspond to waves
outgoing from the bounded medium. Therefore this is the problem of
planar pressure waves in a stratified medium, the direction of
propagation being normal to the interfaces. A generalization to
more complex stratified media would be possible, at least
numerically. In this problem, the Green function corresponds to
the acoustic pressure: of course, it has not the dimension of a
pressure, but the solution for a ``concrete'' problem with source
can be easily solved, as explained in standard textbooks, and
discussed in a recent paper by Levine\cite{levine}.

Other problems correspond to the previous equations:

i) in an approximate way, ignoring higher order duct modes, the
problem of planar waves in a rigid walled duct terminating in two
semi-infinite ducts with different cross sections areas, the
quantities $\zeta $ and $\zeta _{\ell }$ being the ratios of the
areas. The approximation is good at low frequencies.

ii) the problem of a dissipative termination : the terminal
impedances $Z$ and $Z_{\ell }$ can be the impedances of
dissipative media (at low frequencies, a porous medium open to a
large space can be an approximation of a pure resistance, due to
viscous effects).

In all the previous problems, the quantities $\zeta $ and $\zeta
_{\ell }$ are real and positive, the terminations being passive.
For active terminations, they can be negative. An example is the
beginning of self-sustained oscillations in musical instruments: a
nonlinear excitator, like a reed for a clarinet, can be linearized
as a pure resistance. When the main control parameter, i.e. the
pressure in the mouth of the musician, increases, the resistance
becomes negative, the static regime becomes unstable, and an
oscillation starts as an increasing exponential (see e.g. Refs.
\onlinecite{kergo,debut}).

Obviously analogous problems for mechanical vibrations or other
wave fields are numerous.

\section{Closed-form solution for the Fourier Transform\label{compact}}

The Fourier Transform (FT) of $g(x,t)$ is denoted $G(x,\omega )$
(throughout the article, functions of time are written in small
characters, and their FT are written in capital characters). It is
equal to:
\begin{eqnarray}
G(x,\omega )=\int_{-\infty }^{+\infty }g(x,t)e^{-i\omega
t}dt,\text{ where} \label{a1} \\ g(x,t)=\frac{1}{2\pi
}\int_{-\infty }^{+\infty }G(x,\omega )e^{i\omega t}d\omega .
\label{a2}
\end{eqnarray}

The FT of Eqs.~(\ref{z1}) to (\ref{z3}) are found to be:

\begin{eqnarray}
\left( \partial _{xx}^{2}+\omega ^{2}/c^{2}\right) G(x,\omega
)=-\delta (x-x_{0}),  \label{a3}\\ c\zeta \partial _{x}G(x,\omega
)=i\omega G(x,\omega )\;\text{at}\;x=0\text{;} \label{a4} \\
c\zeta _{\ell }\partial _{x}G(x,\omega )=-i\omega G(x,\omega
)\;\text{at}\;x=\ell \text{.} \label{a5}
\end{eqnarray}

While terminal impedances are independent of frequency, boundary
conditions are frequency dependent. Nevertheless a classical,
closed-form, solution is
already known, which has been especially used in Ref. \onlinecite{MF}. If $%
x\neq x_{0}$ solutions of Eq.~(\ref{a3}) can be written as:

\begin{eqnarray*}
G(x,\omega )=A^{-}\cosh \left[ i\omega x/c+\eta \right]
\;\text{if}\;x<x_{0}\text{;}
\\
G(x,\omega )=A^{+}\cosh \left[ i\omega (\ell -x)/c+\eta _{\ell
}\right] \;\text{if}\;x>x_{0}.
\end{eqnarray*}

For the boundary conditions, the following definitions are used:
\begin{eqnarray}
\zeta &=&\coth \eta \;\text{;}\;r=e^{-2\eta }=(\zeta -1)/(\zeta
+1)\, \nonumber \\ \zeta _{\ell } &=&\coth \eta _{\ell }\;\text{;
}\; r_{\ell }=e^{-2\eta _{\ell }}=(\zeta _{\ell }-1)/(\zeta _{\ell
}+1),  \label{11}
\end{eqnarray}
where $r$ and $r_{\ell }$ are the reflection coefficients. The quantity $%
\eta $ satisfies: $2\eta =-\left( \text{ln}\left| r\right| +i\,\text{arg}(r)\right) \text{ }%
[2{\pi }]$. Because $r$ is real, we choose the following
definition:
\begin{equation}
\eta =\eta _{r}+i\mu \pi /2\text{ ; }\mu = 0 \text{ or }1.
\label{J1}
\end{equation}
Two cases exist: i) if $\left| \zeta \right| >1$, $r>0$, $\mu =0;$ ii) if $%
\left| \zeta \right| <1$, $r<0$, $\mu =1.$ Similar remark and
definition can be applied to boundary $x=\ell $:
\begin{equation}
\eta _{\ell }=\eta _{\ell r}+i\mu _{\ell }\pi /2\text{ ; }\mu _{\ell }=0%
\text{ or }1.  \label{J11}
\end{equation}
The case $\zeta =1$ (semi-infinite tube or medium) corresponds to
$\eta =\infty $: it is discussed in the next sections. Except for
the last one, most of the following calculations are valid for all
cases. At $x=x_{0}$, writing the continuity of the function and
the jump of its first derivative, the following result is
obtained:
\begin{equation}
G(x,\omega )=\frac{c}{i\omega }\frac{\cosh \left[ \eta +i\omega x_{0}/c%
\right] \cosh \left[ \eta _{\ell }+i\omega (\ell -x)/c\right]
}{\sinh (i\omega \ell /c+\eta +\eta _{\ell })}\   \label{a7}
\end{equation}

if $x\geq x_{0}$ and a similar result if $x\leq x_{0}$, by
interchanging $x$ and $x_{0}$.

\section{Solution in the time domain (successive reflections)\label%
{successive}}

Eq.~(\ref{a7}) can be transformed in the time domain, leading to a
solution corresponding to the successive reflections of the Green
function in infinite space at the two boundaries. It will be the
reference solution for the check of the validity of the modal
expansion. The $\sinh $ function of the denominator can be written
as

\begin{equation*}
\sinh (i\omega \ell /c+\eta +\eta _{\ell })=\frac{1-e^{-2\eta
-2\eta _{\ell }-2i\omega \ell /c}}{2e^{-\eta -\eta _{\ell
}-i\omega \ell /c}}
\end{equation*}

and, if the modulus of the exponential at the denominator is less
than unity (this is discussed hereafter), as:
\begin{eqnarray}
\sinh ^{-1}(i\omega \ell /c+\eta +\eta _{\ell })\text{ }=2e^{-\eta
-\eta _{\ell }-i\omega \ell /c}  \nonumber \\ \left[ 1+F(\omega
)+F^{2}(\omega )+F^{3}(\omega )+...\right] \text{.} \label{a9}
\end{eqnarray}

$F(\omega )=\exp(-2\eta -2\eta _{\ell }-2i\omega \ell /c)$ is the
function
corresponding to a complete round trip of a wave in the tube, of duration $%
2\ell /c$. Concerning the numerator of (\ref{a7}), it can be written: $%
\exp(+\eta +\eta _{\ell }+i\omega \ell /c)\,G_{p}(x,\omega
)\,c/4$, where:

\begin{eqnarray}
G_{p}(x,\omega )=e^{-i\omega (x-x_{0})/c}+re^{-i\omega
(x+x_{0})/c} \nonumber \\ +r_{\ell }e^{-i\omega (2\ell
-x-x_{0})/c}+rr_{\ell }e^{-i\omega (2\ell -x+x_{0})/c}.
\label{a12}
\end{eqnarray}

Therefore the Green function is:
\begin{equation}
G(x,\omega )=\frac{c}{2i\omega }G_{p}(x,\omega )\left[ 1+F(\omega
)+F^{2}(\omega )+...\right].  \label{a14}
\end{equation}

The factor $G_{p}(x,\omega )/i\omega $ corresponds to the four
``primary'' waves arriving during the first cycle of duration
$2\ell /c$, and this packet is simply reproduced at times $2\ell
/c$, $4\ell /c$, $6\ell /c$, etc... (see for a detailed
explanation e.g. Kergomard \cite{kergo} ). The
inverse FT of the function $G_{p}(x,\omega )/{i\omega }$, denoted $%
h_{p}(x,t) $, is obtained by taking into account the zero
condition for negative times. The result is found to be, whatever
the sign of $\left( x-x_{0}\right) $:
\begin{eqnarray}
h_{p}(x,t)=H\left[ t-\left| x-x_{0}\right| /c\right] +rH\left[ t-(x+x_{0})/c%
\right] +  \nonumber \\ rH\left[ t-(2\ell -x-x_{0})/c\right]
+rr_{\ell }H\left[ t-(2\ell -\left| x-x_{0}\right| )/c\right]
\label{a191}
\end{eqnarray}
where $H(t)$ is the step function. Finally
\begin{eqnarray}
g(x,t)=\frac{c}{2}h_{p}(x,t)\ast \left[ \delta (t)+f(t)+(f\ast f)(t)+...%
\right];  \label{a19} \\ f(t) =rr_{\ell }\delta (t-2\ell/c).
\label{a200}
\end{eqnarray}

Condition of validity of expansion (\ref{a9}) is $\left| rr_{\ell
}\right| <1 $. We notice that if $\zeta $ is real and positive,
$\left| r\right| <1$, and similarly for $\zeta _{\ell }$.
Therefore the condition is satisfied when the two boundaries are
dissipative, or, more precisely, if the
combination of the two reflections is dissipative. The case $%
\left| rr_{\ell }\right| >1$ will be discussed in section \ref
{active}. Other comments can be made:

- the article is limited to purely resistive boundaries, but Eqs.~(\ref{a191}%
) \ and (\ref{a200}) can be generalized to various boundary
conditions defined by a reflection coefficient, $R(\omega )$. This
is done by
replacing the products like $rH(t)$ by the convolution product $(r\ast H)(t)$%
, where $r(t)$ is the inverse FT of  $R(\omega ).$

- for the case under study, we notice that the convolution product
of $n$ times function $f(t)$ is $(rr_{\ell })^{n}\delta (t-2n\ell
/c).$

- if $\zeta $ (respectively $\zeta _{\ell }$) is unity, the
reflection coefficient $r$ (respectively $r_{\ell }$) vanishes, as
well as $f(t)$: the first term of the Green function is the Green
function of an infinite medium, the first two terms correspond to
a semi-infinite medium, etc... As it will be seen in the next
section, no modes can be found for these cases,
because no reflections exist, either $\eta $ or $\eta _{\ell \text{ }}$%
tending to infinity.

- finally, multiplying both members of Eq.~(\ref{a14}) by the
factor $\left[ 1-F(\omega )\right] $, and taking the inverse FT,
it appears that a closed-form exists in the time domain, which is
the basis for the study of the Helmholtz motion of bowed string
instruments (see e.g. Woodhouse\cite {woodhouse} ). It is a
recurrence relationship:
\[
\partial _{t}g(x,t)-rr_{\ell }\partial _{t}g(x,t-2\ell /c)=g_{p}(x,t)\,c/2.
\]

\section{Expansion in resonance modes using the inverse FT \label%
{inverse Fourier}}

Putting expression (\ref{a7}) of the frequency domain in
Eq.~(\ref{a2}) leads to the modal expansion of the time domain
expression. The tool is the residue calculus. If all poles of
expression (\ref{a7}) are simple and located on or above the real
axis, the following equation can be used:
\begin{equation}
g(x,t)=i\Sigma \;\text{ if}\;t>0\;\;\text{and}\;\;0\;\text{
if}\;t<0, \label{a23}
\end{equation}

where $\Sigma $ is the sum of the residues\ of $G(x,\omega )\exp
(i\omega t)$ (see e.g. Morse and Ingard\cite{morse} p 17, changing
$i$ to $-i$).

\subsection{Calculation of the poles\label{lossless poles}}

Zeros of function sinh satisfy:

\begin{equation}
i\omega _{n}= \left[ -\eta -\eta _{\ell }+in\pi \right]c/\ell ,
\label{20ab}
\end{equation}

where $n$ is an integer. In order for the poles to be above the
real axis, the condition is $\eta _{r}+\eta _{\ell r}>0.$ It is
equivalent to the
condition previously obtained for the successive reflections expansion: $%
\left| rr_{\ell }\right| <1.$ Using definition (\ref{J1}),
Eq.~(\ref{20ab}) is rewritten as:
\begin{equation}
\omega _{n}=\left[ n-(\mu +\mu _{\ell })/2\right] \pi c/\ell
+i(\eta _{r}+\eta _{\ell r})c/\ell .  \label{J2}
\end{equation}

As already remarked by several authors, the imaginary part of the
complex frequency is independent of $n,$ and the real part is
independent of the dissipation. Depending on the values of $\zeta
$ and $\zeta _{\ell }$, different cases must be distinguished:

i) if $\left| \zeta \right| >1$ and $\left| \zeta _{\ell }\right| >1$ (real $%
\eta $ and $\eta _{\ell }$): the real part of the frequency
corresponds to the values for pure Neumann conditions (infinite
$\zeta $ and $\zeta _{\ell } $).

ii) if $\left| \zeta \right| >1$ and $\left| \zeta _{\ell }\right|
<1$
(mixed case with either complex $\eta $ or complex $\eta _{\ell }$: either $%
\mu $ or $\mu _{\ell }$ is unity): the real part corresponds to a
problem with different conditions (Neumann and Dirichlet) at $x=0$
at $x=\ell $. The real part of eigenfrequencies is an odd harmonic
of $c/4\ell .$

iii) if $\left| \zeta \right| <1$ and $\left| \zeta _{\ell
}\right| <1$ (complex $\eta $ and $\eta _{\ell }$: $\;\mu =\mu
_{\ell }=1$): the real part corresponds to the values for pure
Dirichlet conditions (zero $\zeta $ and $\zeta _{\ell }$).

Except for case ii), a purely imaginary eigenfrequency exists for
$n=(\mu +\mu _{\ell })/2$.

\subsection{Calculation of the residues\label{lossless residues}}

In all cases, the Taylor expansion of the function $\sinh $ in Eq.~(\ref{a7}%
) at the first order of the quantity $(\omega -\omega _{n}) $ can
be determined. The result is:

\begin{equation}
\sinh \left[ i\omega \ell /c+\eta +\eta _{\ell }\right] \simeq
i(-1)^{n}(\omega -\omega _{n})\ell /c.  \label{23b}
\end{equation}

We get for $\omega $ close to the pole $\omega _{n}$:

\begin{eqnarray}  \label{24c}
G(x,\omega )=-\frac{c^{2}}{\omega _{n}\ell }\frac{f_{n}(x)f_{n}(x_{0}) }{%
(\omega -\omega _{n})}; \\ f_{n}(x)=\cosh (\eta +i\omega _{n}x/c)
\label{299}
\end{eqnarray}

or $f_{n}(x)=(-1)^{n}\cosh (i\omega _{n}(\ell -x)/c+\eta _{\ell
})$. The residue corresponding to the pole $\omega =0$, remains to
be calculated. For small $\omega $,
\begin{equation}
G(x,\omega )=\frac{c}{i\omega }\frac{\cosh \eta \cosh \eta _{\ell
}}{\sinh (\eta +\eta _{\ell })}=\frac{c}{i\omega }\frac{1}{\zeta
^{-1}+\zeta _{\ell }^{-1}}.  \label{24d}
\end{equation}

Using Eq.~(\ref{a23}), the inverse FT of $G(x,\omega ) $ is
obtained:

\begin{equation}
g(x,t)=H(t)\frac{c^{2}}{\ell }%
\mathrel{\mathop{\sum }\limits_{n}}%
\frac{f_{n}(x)f_{n}(x_{0})}{i\omega _{n}}e^{i\omega _{n}t}+\frac{cH(t)}{%
\zeta ^{-1}+\zeta _{\ell }^{-1}}.  \label{29}
\end{equation}

Some comments can be made:

- the formula is valid for all aforementioned cases;

- the mode shapes $f_{n}(x)$ are complex-valued functions of the
space variable, meaning that the shape is varying with time (for a
discussion on complex modes, see e.g.Ref. \onlinecite{arruda}).
The question of their orthogonality will be discussed in section
\ref{separation}. Notice that
functions $f_{n}(x)$ do not fulfill the same boundary conditions than $%
G(x,\omega )$: the boundary conditions are (\ref{a4}) and
(\ref{a5}), but where $\omega $ is replaced by $\omega _{n};$

- the last term in Eq. (\ref{29}) is a constant mode. If one of
the impedances $\zeta $ or $\zeta _{\ell }$ is zero, it
disappears, as it is intuitive, in order to satisfy a Dirichlet
condition. It is a trivial solution of the wave equation and the
boundary conditions, and can be
compared to the DC component of a periodic signal. When both $\zeta $ and $%
\zeta _{\ell }$ tend to infinity, the boundaries tend to Neumann
boundaries, and the combination of the non oscillatory mode of
frequency $\omega _{0}$ and the constant mode results in a uniform
(i.e. constant in space) mode
which increases linearly with time. The calculation is done as follows: if $%
\eta $ and $\eta _{\ell }$ tend to zero, $\omega _{0}$ tends to
zero, and the factor $\exp (i\omega _{0}t)$ can be written as:
$1+i\omega _{0}t.$ The zeroth order term is equal to the opposite
of the constant mode, and only the linear term remains. The result
is $H(t)\,tc^{2}/\ell $, and its FT is $-c^{2}/\ell \omega ^{2}$.
This mode is the classical uniform mode existing for instance in
3D cavities with rigid walls: curiously it exists in the standard
textbooks (see e.g. Ref.\onlinecite{morse}, page 571), but the
time domain expression is not given. This mode is similar to the
well known planar guided mode, existing in ducts with rigid walls,
whatever the geometrical shape.

- the imaginary part of the complex frequencies being independent
of $n$, the decay is identical for all non constant modes;

- for the above-considered case i), we notice that $\omega
_{-n}=-\omega _{n}^{\ast }$ and $f_{-n}(x)=f_{n}^{\ast }(x),$ and,
more generally:
\begin{eqnarray}
i\omega _{\nu }&=&(i\omega _{n})^{\ast }\; \text{ ; } \; f_{\nu
}(x)=(-1)^{\mu }f_{n}^{\ast }(x)\,, \label{J3} \\ \text{where }
\nu &=&-n+\mu +\mu _{\ell } \text{ is an integer.}\label{J3nu}
\end{eqnarray}
As a consequence, the solution $g(x,t)$ is real. It could be
possible to transform the sum by adding the two oscillating terms
corresponding to $n$ and $\nu $, when $n\neq \nu ,$ as it is
usually done for non dissipative boundaries. Nevertheless it
appears that the formulas become intricate.

- Eqs.~(\ref{24c}) and (\ref{24d}) lead directly to another form
of the FT of result (\ref{29}), written as a series:

\begin{equation}
G(x,\omega )=-\frac{c^{2}}{\ell }%
\mathrel{\mathop{\sum }\limits_{n}}%
\frac{f_{n}(x)f_{n}(x_{0})}{\omega _{n}(\omega -\omega _{n})}+\frac{c}{%
i\omega }\frac{1}{\zeta ^{-1}+\zeta _{\ell }^{-1}}.  \label{3000}
\end{equation}

\begin{figure}

\ifgalleyfig   
  \includegraphics[width=3.25in]{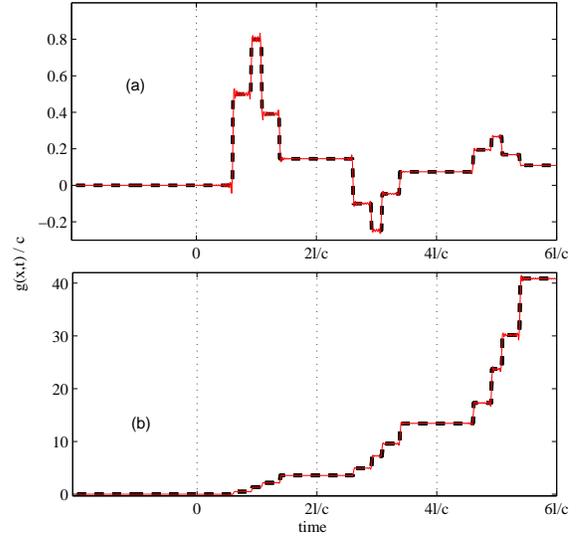}
\else
  \ifoutputfig  
  \includegraphics[width=5.5in]{fig2.eps}
  \fi
\fi

\caption [ Normalized Green function as a function of time:
comparison between the successive reflections method (dotted line)
and modal expansion (solid line, $102$ modes, i.e. maximum
$n=50$). Locations of the source and receiver are $x_{0}/\ell
=0.15$ and $x/\ell =0.76,$ respectively.\newline a) passive
boundaries $\zeta =4.$ ; $\zeta _{\ell }=0.1$ (constant mode
$=0.097$).  b) active boundary $\zeta =\zeta _{\ell }=-4.$
(constant mode$=-2$).] {\label{fig2} \ifgalleyfig {Normalized
Green function as a function of time: comparison between the
successive reflections method (dotted line) and modal expansion
(solid line, $102$ modes, i.e. maximum $n=50$). Locations of the
source and receiver are $x_{0}/\ell =0.15$ and $x/\ell =0.76,$
respectively.\newline (a) passive boundaries $\zeta =4.$ ; $\zeta
_{\ell }=0.1$ (constant mode$=0.097$).  (b) active boundary $\zeta
=\zeta _{\ell }=-4.$ (constant mode$=-2$).} \fi}
\end{figure}

An example of comparison of the successive reflections method and
modal expansion is shown in figure \ref{fig2}(a), for conditions
close to Neumann and Dirichlet. We notice that it is satisfactory.
The Gibbs phenomenon appears, because of the truncated series of
modes, ensuring the correct accordance between the two methods.
Moreover this accordance confirms the existence of the constant
mode. Decreasing of the maxima is due to the dissipation at the
boundaries: for pure Neumann and Dirichlet conditions, the shape
would be similar, but perfectly periodical.

\subsection{The case of active boundaries\label{active}}

What happens when the combination of boundaries is active, i.e.
when $\left| rr_{\ell }\right| >1,$ or $\eta _{r}+\eta _{\ell
r}<0$ (at least one of the impedances $\zeta $ or $\zeta _{\ell }$
is negative)? It is possible to prove that Eqs.~(\ref{a19}) and
(\ref{29}){\bf \ }{\it remain valid for
active boundary conditions, }as explained hereafter. The real part of $%
i\omega _{n}$ being independent of $n$, this suggests to use a new function $%
\widetilde{g}(x,t)=g(x,t)\exp (-\widetilde{\eta }t),$ where
$\widetilde{\eta }>-\eta _{r}-\eta _{\ell r}>0$, which can be
substituted in the initial problem in order for the poles to be
located again on or above the real axis. Eq.~(\ref{z1}) becomes
\[
\partial _{xx}^{2}\widetilde{g}(x,t)-c^{-2}\left[ \partial _{t}+\widetilde{%
\eta }\right] ^{2}\widetilde{g}(x,t)=-\delta (x-x_{0})e^{-\widetilde{\eta }%
t}\delta (t)
\]
and similarly for Eqs.~(\ref{z2}) and (\ref{z3}). It is equivalent
to use an
appropriate Laplace Transform. Going in the frequency domain leads to Eqs.~(%
\ref{a3} to \ref{a5}), where $G(x,\omega )$ is replaced by $\widetilde{G}%
(x,\omega )$ and $i\omega $ by ($i\omega +\widetilde{\eta })$, and
a similar result for Eq.~(\ref{a7}). The analysis of both
successive reflections and
poles and residues leads to the result $\widetilde{g}(x,t)=g(x,t)\exp (-%
\widetilde{\eta }t)$, where $g(x,t)$ is given by Eqs.~(\ref{a19})
and (\ref {29}), respectively, and the proof is achieved. We do
not repeat here the complete procedure. We notice that for the
case $\eta _{r}+\eta _{\ell r}=0$, one boundary is active and the
other one is passive: eigenfrequencies $\omega _{n}$ are real
while modes are complex. Figure \ref{fig2}(b) shows an example of
result.

\section{ Method of separation of variables\label{separation}}

\subsection{Second order homogeneous equation with initial conditions\label%
{sep2nd}}

\subsubsection{Derivation of the modes}

Oliveto and Santini \cite{oliveto}, and Guyader \cite{guyader} have treated a particular case of the problem (zero $%
\zeta $ , large $\zeta _{\ell }$) using the method of separation
of
variables. He gets non orthogonal modes for the common scalar product $%
\int_{0}^{\ell }f_{n}(x)f_{m}(x)dx$. We will see that the method
is valid whatever the values of the two boundary conditions, and
how the derivation can be simplified.

We are searching for solutions $p(x,t)$ of homogeneous equation
(\ref{z1}) (without second member), with boundary conditions
(\ref{z2}) and (\ref{z3}), and with given initial conditions.
Assuming that the general solution is a superposition of solutions
with separate variables, the solutions with separate variables are
written in the following form:
\begin{eqnarray}
p(x,t)=f(x)h(t);  \label{30} \\ h(t)=B^{+}e^{i\omega
t}+B^{-}e^{-i\omega t};  \label{31} \\ f(x)=\cosh (i\omega
x/c+\varphi )\text{.}  \label{32}
\end{eqnarray}

Decomposition (\ref{30}) differs from the ordinary FT, because a priori $%
\omega $ is a complex quantity, depending on the boundary
conditions. Considering first the solution $B^{+}e^{i\omega t}$,
this leads to :

\begin{eqnarray}
\zeta \omega \sinh \varphi =\omega \cosh \varphi;  \label{32A} \\
\omega \zeta _{\ell }\sinh (i\omega \ell /c+\varphi )=-\omega
\cosh (i\omega \ell /c+\varphi ).  \label{32B}
\end{eqnarray}

$\omega =0$ is a solution, corresponding to the constant mode. The
other modes are given by Eq.~(\ref{32A}): $\sinh (\varphi -\eta
)=0$, thus:
\begin{equation}
f(x)=\cosh (i\omega x/c+\eta )\text{.}  \label{32C}
\end{equation}

Actually there is a sign $\pm $ in the right-hand side member of
Eq.~( \ref {32C}), but it is without importance, because it can be
included in the coefficient $B^{+}$ of the solution. The
eigenvalues equation is deduced from Eqs.~(\ref{32A}) and
(\ref{32B}), as follows:
\begin{equation}
\sinh (i\omega \ell /c+\eta +\eta _{\ell })=0,  \label{32D}
\end{equation}

the solutions being given by (\ref{20ab}). The solution in time $%
B^{-}e^{-i\omega t}$ does not lead to new solutions for $f(x)$,
therefore, assuming the solutions form a basis of solutions (this
is discussed in section \ref{ortho}) , the general solution of a
problem with initial conditions can be written as:

\begin{equation}
p(x,t)=
\mathrel{\mathop{\sum }\limits_{n}}%
A_{n}f_{n}(x)e^{i\omega _{n}t}+A,  \label{35}
\end{equation}

where $\omega _{n}$ and $f_{n}(x)$ are given by Eqs.~(\ref{20ab})
and (\ref {299}), respectively, and the coefficients $A_{n}$ and
$A$ depend on the initial conditions, and can be determined using
the orthogonality relation of the modes. $A$ is the coefficient of
the constant mode.

\subsubsection{Orthogonality relationship between the modes: first approach}

In order to derive an orthogonality relationship between the modes
the common product is first calculated:
\begin{equation}
{\it \Lambda }_{nm}=\int_{0}^{\ell }f_{n}(x)f_{m}(x)dx.
\label{35a}
\end{equation}

Because $f_{n}(x)=(-1)^{\mu}f_{\nu}^{\ast }(x)$, the calculation
of the quantities defined in Eq.~(\ref{35a}) for all values of the
index $n$ is
equivalent to the calculation of the quantities defined when replacing $%
f_{m}(x)$ by its conjugate. Writing
\begin{eqnarray*}
&&\int_{0}^{\ell }\left[ f_{n}(x)\frac{d^{2}f_{m}(x)}{dx^{2}}-f_{m}(x)\frac{%
d^{2}f_{n}(x)}{dx^{2}}\right] dx= \\
&&\left[ f_{n}(x)\frac{df_{m}(x)}{dx}-f_{m}(x)\frac{d^{2}f_{n}(x)}{dx^{2}}%
\right] _{0}^{\ell }
\end{eqnarray*}


and using Eq.~(\ref{299}), the following result is obtained:
\[
\frac{(\omega _{m}^{2}-\omega _{n}^{2})}{c^{2}}{\it \Lambda
}_{nm}=i(\omega _{m}-\omega _{n})\left[
\frac{f_{n}(0)f_{m}(0)}{\zeta }+\frac{f_{n}(\ell )f_{m}(\ell
)}{\zeta _{\ell }}\right] .
\]

For $\omega _{m}\neq \omega _{n}$, because $\omega _{m}+\omega
_{n}\neq 0$, the expression of ${\it \Lambda }_{nm}$ is deduced.
For $\omega _{m}=\omega _{n},$ the calculation is straightforward.
The general formula is found to be:
\begin{equation}
{\it \Lambda }_{nm}=\frac{ci}{\omega _{m}+\omega _{n}}\left[ \frac{%
f_{n}(0)f_{m}(0)}{\zeta }+\frac{f_{n}(\ell )f_{m}(\ell )}{\zeta _{\ell }}%
\right] +\frac{1}{2}\ell \delta _{nm}  \label{433}
\end{equation}

where $\delta _{nm}$ is the Kronecker symbol, or:
\begin{equation}
{\it \Lambda }_{nm}=-\frac{c}{2}\frac{\sinh 2\eta +(-1)^{n+m}\sinh
2\eta _{\ell }}{i(\omega _{m}+\omega _{n})}+\frac{1}{2}\ell \delta
_{nm}. \label{43}
\end{equation}

Modes are found to be non orthogonal for the product defined by
(\ref{35a}), but, as shown by Guyader\cite{guyader}, it is
possible to solve the problem
from the knowledge of initial conditions. When dissipation tends to zero ($%
\eta_{r} $ and $\eta _{{\ell }r}$ tend to zero), the first term
does \ not vanish, tending to $(-1)^{\mu}\frac{1}{2}\ell \delta
_{n,{\nu}}$. This is
due to the choice of considering separately the modes $\omega _{n}$ and $%
\omega _{\nu}.$

Otherwise formula (\ref{433}) remains valid when one of the modes
is the constant mode $f(x)=1$, and the other one a non constant
mode:

\begin{eqnarray}
{\it \Lambda }_{n}=\int_{0}^{\ell }f_{n}(x)dx=-\frac{c}{i\omega
_{n}}\left[ \frac{f_{n}(0)}{\zeta }+\frac{f_{n}(\ell )}{\zeta
_{\ell }}\right]  \nonumber
\\
=-\frac{c}{i\omega _{n}}(\sinh \eta +(-1)^{n}\sinh \eta _{\ell }).
\label{43bb}
\end{eqnarray}

Finally the product of the constant mode by itself is $\ell .$

\subsubsection{Solution with respect to initial conditions \label{initial}}

According to Eq.~(\ref{35}), the initial conditions are:
\begin{eqnarray}
p(x,0)=%
\mathrel{\mathop{\sum }\limits_{n}}%
A_{n}\cosh (i\omega _{n}x/c+\eta )+A\text{;}  \label{44} \\
\partial _{t}p(x,0)=%
\mathrel{\mathop{\sum }\limits_{n}}%
A_{n}i\omega _{n}\cosh (i\omega _{n}x/c+\eta ).  \label{45}
\end{eqnarray}

Using Eq.~(\ref{43}) for a non constant mode $m$, \ the following
results are obtained:
\begin{eqnarray}
\int_{0}^{\ell }p(x,0)f_{m}(x)dx=%
\mathrel{\mathop{\sum }\limits_{n}}%
A_{n}{\it \Lambda }_{nm}+A{\it \Lambda }_{m};  \label{46} \\
\int_{0}^{\ell }\partial _{t}p(x,0)f_{m}(x)dx=%
\mathrel{\mathop{\sum }\limits_{n}}%
A_{n}i\omega _{n}{\it \Lambda }_{nm}.\text{ }  \label{48}
\end{eqnarray}

Multiplying Eq.~(\ref{46}) by $i\omega _{m\text{ }},$ then adding
Eq.~(\ref {48}), leads to:
\begin{eqnarray}
\int_{0}^{\ell }\left[ i\omega _{m}p(x,0)+\partial
_{t}p(x,0)\right] f_{m}(x)dx  \nonumber \\
=i%
\mathrel{\mathop{\sum }\limits_{n}}%
A_{n}(\omega _{m}+\omega _{n}){\it \Lambda }_{nm}+iA\omega _{m}{\it \Lambda }%
_{m}  \nonumber \\
=-c%
\mathrel{\mathop{\widehat{\sum }}\limits_{n}}%
A_{n}\left[ \frac{f_{n}(0)f_{m}(0)}{\zeta }+\frac{f_{n}(\ell )f_{m}(\ell )}{%
\zeta _{\ell }}\right] +iA_{m}\ell \omega _{m}  \label{497} \\
=-c\left[ \frac{f_{m}(0)p(0,0)}{\zeta }+\frac{f_{m}(\ell )p(\ell
,0)}{\zeta _{\ell }}\right] +iA_{m}\ell \omega _{m}.  \label{498}
\end{eqnarray}

Notation $\widehat{\sum }$ for the series in Eq.~(\ref{497})
indicates that it involves the constant mode. As noticed by
Guyader\cite{guyader}, this
series is related to the initial conditions at the two ends $x=0$ and $%
x=\ell $. Thus for a non constant mode:
\begin{eqnarray}
A_{n}\ell i\omega _{n}=\int_{0}^{\ell }\left[ i\omega
_{n}p(x,0)+\partial _{t}p(x,0)\right] f_{n}(x)dx  \nonumber \\
+cp(0,0)\sinh \eta +cp(\ell ,0)\varepsilon _{n}\sinh \eta _{\ell
}. \label{49}
\end{eqnarray}
The following property is deduced from Eq.~(\ref{J3}): $A_{\nu}f_{%
\nu}(x)=A_{n}^{\ast }f_{n}^{\ast }(x)$, thus $p(x,t)$ is real. Calculating $%
\int_{0}^{\ell }\partial _{t}p(x,0)dx$, we similarly get
coefficient $A$:

\begin{equation}
A=\frac{c^{-1}\int_{0}^{\ell }\partial _{t}p(x,0)dx+p(0,0)\tanh
\eta +p(\ell ,0)\tanh \eta _{\ell }}{\tanh \eta +\tanh \eta _{\ell
}}.  \label{50a}
\end{equation}

What is the condition for which this coefficient vanishes? If for
instance at $x=0$,$\ \ \zeta $ is zero, $\eta $ is infinite, and,
according to the boundary condition, $p(0,0)$ vanishes, thus $A$
vanishes too. This confirms the remark concerning result
(\ref{29}).\newline Using the initial conditions for the Green
function found in (\ref{a19}), it is possible to check result
(\ref{29}), but this will be done hereafter using the equation
with source.

\subsection{ First order system of equations, orthogonality and completeness
of the modes \label{ortho}}

\subsubsection{ Introduction}

In this section we will prove that the modes form a Riesz basis in
the space of solutions of a closely related problem, and give the
expression of a scalar product making the modes orthogonal. As an
introduction we show that a modified scalar product leads to the
orthogonality of modes, except the constant one. For vibrating
systems, the product defined by (\ref{35a}) corresponds to the
product with respect to the mass, a complement being the
calculation of the product related to the stiffness (see e.g.
Meirovitch \cite{meiro} ):
\begin{equation*}
{\it \Lambda }_{nm}^{\prime }=\int_{0}^{\ell }\frac{d}{dx}f_{n}(x)\frac{d}{dx%
}f_{m}(x)dx.
\end{equation*}

By integrating by parts, and using Eq.~(\ref{433}), this product,
for $n\neq m$, is found to be equal to:
\begin{equation*}
{\it \Lambda }_{nm}^{\prime }=\left[ f_{n}(x)d_{x}f_{m}(x)\right]
_{0}^{\ell }+\frac{\omega _{m}^{2}}{c^{2}}{\it \Lambda
}_{nm}=-\frac{\omega _{n}\omega _{m}}{c^{2}}{\it \Lambda }_{nm}.
\end{equation*}

Therefore the modes become orthogonal if we define a new product,
as follows:
\begin{equation}
\int_{0}^{\ell }\left[ \partial _{x}p_{n}\partial _{x}p_{m}-\frac{1}{c^{2}}%
\partial _{t}p_{n}\partial _{t}p_{m}\right] _{t=0}dx=\delta _{nm}\ell \omega
_{n}^{2}/c^{2},  \label{K3}
\end{equation}
where $p_{n}=p_{n}(x,t)=f_{n}(x)\exp (i\omega _{n}t)$ and
similarly for index $m$. We remark that the modes $p_{n}$ and
$p_{\nu }=(-1)^{\mu }p_{n}^{\ast }$ are orthogonal for this
product. For the calculation of the solution from initial
conditions, using Eq.~(\ref{35}) at $t=0$, the following result is
obtained:
\begin{equation}
\int_{0}^{\ell }\left[ \frac{d}{dx}f_{n}(x)\partial
_{x}p(x,0)-\frac{i\omega
_{n}}{c^{2}}f_{n}(x)\partial _{t}p(x,0)\right] dx=A_{n}\frac{\omega _{n}^{2}%
}{c^{2}}\ell .  \label{K4}
\end{equation}

As a consequence, the initial conditions need to be written by
using the derivatives of the function $p(x,t)$ with respect to
abscissa and time, respectively. Result (\ref{49}) can be checked
by integrating by parts the first term of the integral.
Nevertheless, the product (\ref{K3}) is not useful for the
constant mode, and the first method needs to be used (see
subsection \ref{initial}). Moreover this derivation does not prove
that the product is a scalar product, and that the modes form a
basis of the space of solutions of the problem. This will be done
hereafter.

\subsubsection{Riesz basis of the modes}

Several works have been done by mathematicians concerning spectral
operators when boundary conditions are not simple conditions like
Neumann or Dirichlet conditions. We quote the work by
Russell\cite{Rus}, Majda\cite{Majda}, Lagnese\cite{lagn}, Banks et
al\cite{banks}, Darmawijoyo and Van Horssen\cite {DVD}, Cox and
Zuazua \cite{cox}. Rideau \cite{rideau} has treated the 1D case
with a (unique) resistive termination, giving explicitly a scalar
product (see also Ref.%
\onlinecite{int}). We generalize his calculation using a similar
method, by considering the wave equation with source in the
following form:
\begin{equation}
\partial _{t}{\bm \psi }(x,t)={\bm A}{\bm \psi }(x,t)+{\bm \phi }_{s}(x,t),%
\text{ }  \label{A20}
\end{equation}
where ${\bm \psi }(x,t)=(p,v)^{T}$, $p$ and $v/(\rho c)$ being the
acoustic pressure and velocity, respectively. Operator \ ${\bm A}$
is:

\begin{equation}
\text{\ }{\bm A}=\left(
\begin{array}{cc}
0 & -c\partial _{x} \\ -c\partial _{x} & 0
\end{array}
\right) ,  \label{A3}
\end{equation}
and boundary conditions are written as:
\begin{equation}
p(0,t)=-\zeta v(0,t)\text{ \ and \ }p(\ell ,t)=\zeta _{\ell
}v(\ell ,t)\text{ \ }\forall t.  \label{A2}
\end{equation}

The family of eigenelements of ${\bm A}$ are found to satisfy:
\begin{equation}
\lambda _{n}p_{n}(x)=-c\partial _{x}v_{n}(x)\;;\;\lambda
_{n}v_{n}(x)=-c\partial _{x}p_{n}(x),\;  \label{A41}
\end{equation}
thus
\begin{eqnarray}
\left(
\begin{array}{c}
p_{n}(x) \\ v_{n}(x)
\end{array}
\right)  &=&\left(
\begin{array}{c}
\cosh (\lambda _{n}x/c+\eta ) \\ -\sinh (\lambda _{n}x/c+\eta )
\end{array}
\right)   \label{A5} \\ \lambda _{n} &=&\left( -\eta -\eta _{\ell
}+in\pi \right) c/\ell =i\omega _{n}  \label{A6}
\end{eqnarray}
(see Eq.~(\ref{20ab})). $p_{n}(x)=f_{n}(x)$ and $\lambda _{n}$ are
identical to the eigenfunctions and eigenvalues found before.
Nevertheless the constant mode is eliminated (except for the very
particular case $\eta =-\eta _{\ell }$), because the boundary
conditions are slightly different: Eqs.~(\ref{z2}) and (\ref{z3})
are obtained by deriving Eqs.~(\ref{A2}) with respect to $t$. In
Eq.~(\ref{A5}) the argument of the hyperbolic functions can be
written as:
\begin{eqnarray}
\lambda _{n}x/c+\eta  &=&\alpha (x)+i\beta _{n}(x);  \label{A7} \\
\alpha (x) &=&-\eta _{\ell r}x/\ell +\eta _{r}\left( 1-x/\ell
\right) ; \label{A8} \\ \beta _{n}(x) &=&\pi \left[ -\mu _{\ell
}x/\ell +\mu \left( 1-x/\ell \right) \right] /2 +n\pi x/\ell.
\label{A9}
\end{eqnarray}
(see definitions (\ref{J1}) and (\ref{J11})). Denoting ${\bm \psi }%
_{n}^{\alpha }(x)=(p_{n}(x),v_{n}(x))^{T}$, we show in Appendix A
that the family of elements ${\bm \psi }_{n}^{\alpha }(x)$ is a
Riesz basis, i.e. a complete basis of  elements, which become
orthogonal
 for the following scalar product:
\begin{equation}
<{\bm \psi },{\bm \varphi }>_{H}^{\alpha }=\int_{0}^{\ell }{\bm \varphi }%
^{T\ast }{\bm G}_{2\alpha }(x){\bm \psi }dx \,, \label{A13}
\end{equation}

\begin{equation}
\text{where  } {\bm G}_{\alpha }(x)=\left(
\begin{array}{cc}
\cosh \alpha (x) & \sinh \alpha (x) \\ \sinh \alpha (x) & \cosh
\alpha (x)
\end{array}
\right) .  \label{A12}
\end{equation}

For a given vector ${\bm \psi }=(p,v)^{T}$, the value of the
scalar product with eigenvector ${\bm \psi }_{n}^{\alpha }$ is
found to be:
\begin{equation}
<{\bm \psi },{\bm \psi }_{n}^{\alpha }>_{H}^{\alpha
}=\int_{0}^{\ell }\left[ p(x,t)p_{n}(x)-v(x,t)v_{n}(x)\right] dx.
\label{A18}
\end{equation}

This is in accordance with the product (\ref{K3}). A direct
application of
this result is the solution of Eq.~(\ref{A20}) with initial conditions ${\bm %
\psi }(x,0)=(p(x,0),v(x,0))^{T}$. The modal decomposition is
uniquely
determined as ${\bm \psi }(x,t)=%
\mathrel{\mathop{\sum }\limits_{n}}%
h_{n}(t){\bm \psi }_{n}^{\alpha }(x)$ in the energy space $H$, and
leads to the following family of {\em decoupled} ordinary
differential equations:
\begin{eqnarray}
\ell \left[ \partial _{t}h_{n}-\lambda _{n}h_{n}\right]  &=&<{\bm \phi }%
_{s}(x,t),{\bm \psi }_{n}^{\alpha }>_{H}^{\alpha };  \label{A21}
\\ h_{n}(0) &=&<{\bm \psi }(x,0),{\bm \psi }_{n}^{\alpha
}>_{H}^{\alpha }. \label{A211}
\end{eqnarray}

This result can be first applied to the calculation done in
section \ref {sep2nd}. In order to find a solution $\chi (x,t)$ of
the second order equation without source, we denote ${\bm \psi
}(x,t)=(\partial _{t}\chi
(x,t),-c\partial _{x}\chi (x,t))^{T}$, and obtain by integrating ${\bm \psi }%
(x,t)$ with respect to $t:$
\begin{equation}
\chi (x,0)=%
\mathrel{\mathop{\sum }\limits_{n}}%
\lambda _{n}^{-1}h_{n}(0)p_{n}(x)+A.  \label{A19}
\end{equation}

Using Eqs.~(\ref{A211}), (\ref{A18}), and replacing $\chi (x,t)$ by $p(x,t)$%
, $p_{n}(x)$ by $f_{n}(x),$ $\lambda _{n}$ by $i\omega _{n}$, and
\ $\lambda _{n}^{-1}h_{n}(0)$ by $A_{n}$, formula (\ref{K4}) is
checked. Notice that coefficient $A$ cannot be directly determined
with this method.

\subsubsection{Example of the Green function}

Similarly, the Green function can be calculated by using the
previous result. In order for the unknown function to satisfy the
boundary conditions (\ref{A2}), or (\ref{z2}) and (\ref{z3}), it
is convenient to define the following vectors:
\begin{equation}
{\bm \psi }(x,t)=\left(
\begin{array}{c}
\partial _{t}g(x,t) \\
-c\partial _{x}g(x,t)
\end{array}
\right) \text{; }{\bm \phi }_{s}(x,t)=\left(
\begin{array}{c}
c^{2}\delta (x-x_{0})\delta (t) \\ 0
\end{array}
\right).  \label{A221}
\end{equation}

The first row of Eq.~(\ref{A20}) is Eq.~(\ref{z1}), while the
second one comes from the definition of vector ${\bm \psi }$.
Using Eq.~(\ref{A21}),
the solution is found to be: ${\bm \psi }(x,t)=%
\mathrel{\mathop{\sum }\limits_{n}}%
{\bm \psi }_{n}^{\alpha }(x)h_{n}(t)$, where
\begin{equation}
\partial _{t}h_{n}-\lambda _{n}h_{n}=c^{2}\ell ^{-1}p_{n}(x_{0})\delta (t).
\label{A24}
\end{equation}

The initial conditions for the Green function imply ${\bm \psi
}(x,t)=0$ for $t<0$, therefore $h_{n}(t)=0$ for $t<0$ $.$ Thus the
solution of Eq.~(\ref {A24}) is:
\[
h_{n}(t)=A_{n}e^{\lambda _{n}t}H(t) \text{; }A_{n}=c^{2}\ell
^{-1}p_{n}(x_{0}).
\]

As a consequence,
\begin{equation}
\left(
\begin{array}{c}
\partial _{t}g(x,t) \\
-c\partial _{x}g(x,t)
\end{array}
\right) =%
\mathrel{\mathop{\sum }\limits_{n}}%
A_{n}\left(
\begin{array}{c}
p_{n}(x) \\ v_{n}(x)
\end{array}
\right) e^{\lambda _{n}t}H(t).  \label{A25}
\end{equation}

Integrating the first row with respect to time leads to:
\begin{equation}
g(x,t)=H(t)%
\mathrel{\mathop{\sum }\limits_{n}}%
\left[A_{n}p_{n}(x)e^{\lambda _{n}t}+A(x)\right].  \label{A26}
\end{equation}

Derivating this expression with respect to $x$ and using the second row of (%
\ref{A25}) leads to $\partial _{x}A(x)=0$ , thus $A$ is a
constant, as expected. In order to deduce the value of this
constant, we need the following result:
\begin{equation}
\partial _{t}g(x,0)=p(x,0)=c^{2}\delta (x-x_{0}).  \label{A27}
\end{equation}
It is obtained by derivating the first row of (\ref{A25}) with
respect to time, and the second row of (\ref{A25}) with respect to
abscissa, leading to $p(x,0)\delta (t)=c^{2}\delta (x-x_{0})\delta
(t)$ \ (remind that $\partial _{t}\left[ F(t)H(t)\right]
=H(t)\partial _{t}F(t)+F(0)\delta (t)$ ). The end of the
calculation is done in section \ref{initial}, giving
Eq.~(\ref{50a}),
by replacing $p(x,0)$ by $g(x,0)$ and taking into account that $%
g(0,0)=g(\ell ,0)=0.$ We notice that the calculation is valid for
both passive and active boundaries.

\section{ Eigenmodes expansion in the frequency domain: biorthogonality\label%
{eigenfrequencies}}

Frequency domain approach is very popular in acoustics (see e.g. Ref.%
\onlinecite{morse} ), and leads to the use of biorthogonality (see e.g. Ref.%
\onlinecite{MF}, p.884) of modes, except when the boundary
impedances are imaginary, corresponding to non dissipative
boundaries: for that case, modes are orthogonal, and the laplacian
operator is self-adjoint. In this section we limit the discussion
to the Green function calculation, and use successively the two
above-used approaches : the second order equation, then the system
of two first order equations, ignoring the constant mode. Because
we are now in the Fourier domain, equations are ordinary
differential equations, biorthogonality theory ensuring the
completeness of the modes family.

\subsection{Solution of the second order equation}

\subsubsection{Modal expansion}

In order to calculate the inverse FT of $G(x,\omega ),$ another
solution is possible: the expansion of\ $G(x,\omega )$ in
eigenmodes. This is done for a particular case by Filippi
\cite{filippi} (p. 58 : this author considers another type of
excitation instead of the Dirac function, thus uses the Laplace
Transform instead of the FT; notice that the constant mode is
missing in this work). We will see how this method leads to the
same poles and residues that the direct method using the
closed-form expression (\ref {a7}). We are searching for the
following expansion :

\begin{equation}
G(x,\omega )=%
\mathrel{\mathop{\sum }\limits_{n}}%
G_{n}(x,\omega ),  \label{60c}
\end{equation}

where the eigenmodes $G_{n}(x,\omega )$ are solutions of:
\begin{equation}
\left[ \partial _{xx}^{2}+\theta _{n}^{2}(\omega )/c^{2}\right]
G_{n}(x,\omega )=0,  \label{60d}
\end{equation}

and satisfy the boundary conditions (\ref{a4}) and (\ref{a5}). The
key point is that eigenmodes $G_{n}(x,\omega )$ and
eigenfrequencies $\theta _{n}(\omega )$ depend on frequency
$\omega:$ this is due to the boundary conditions, which are of
Robin type. Solutions of Eqs.~(\ref{60d}) can be written as
follows:
\begin{equation}
G_{n}(x,\omega )=\cosh (i\theta _{n}(\omega )x/c+\varphi
_{n}(\omega )) \label{61}
\end{equation}

where $\theta _{n}(\omega )$ and $\varphi _{n}(\omega )$ are given
by the boundary conditions. Thus they satisfy:
\begin{eqnarray}
\theta _{n}(\omega )\tanh \varphi _{n}(\omega )=\omega/ \zeta
\text{; } \label{62} \\ \theta _{n}(\omega )\tanh (i\theta
_{n}(\omega )\ell /c+\varphi _{n}(\omega ))=-\omega/\zeta _{\ell
}.  \label{63}
\end{eqnarray}

Eliminating quantity $\varphi _{n}(\omega )$, the eigenvalues are
found to satisfy the following equation:
\begin{equation}
\tanh (i\theta _{n}(\omega )\ell /c)\left[ \theta _{n}(\omega
)+\frac{\omega
^{2}}{\theta _{n}(\omega )\zeta \zeta _{\ell }}\right] =-\omega \left[ \frac{%
1}{\zeta }+\frac{1}{\zeta _{\ell }}\right] .  \label{63a}
\end{equation}

When $\theta _{n}(\omega )$ and $\omega $ are not simultaneously
zero, this equation can be rewritten as:
\begin{equation}
e^{2i\theta _{n}\ell /c}=\left[ \frac{\theta _{n}\zeta -\omega
}{\theta
_{n}\zeta +\omega }\right] \left[ \frac{\theta _{n}\zeta _{\ell }-\omega }{%
\theta _{n}\zeta _{\ell }+\omega }\right] \,.  \label{64}
\end{equation}

Calculation of all solutions of this equation is not necessary,
only one of them being useful in the following. Operator
$D=\partial _{xx}^{2}$ is formally equal to its adjoint
$\overline{D}$, but the boundary conditions are different
(conditions for $\overline{D}$ are complex conjugate of conditions
for $D$). Modes of $\overline{D}$ are the complex conjugate of
modes $G_{n}(x,\omega )$ (they are equal to modes $G_{n}(x,\omega
)$ only if $\zeta $ and $\zeta _{\ell }$ are imaginary, because of
the factor $i$ in boundary conditions (\ref{a4}) and (\ref{a5})).
Thus in general operator $D$
is not self-adjoint, and eigenmodes $G_{n}(x,\omega )$ and $\overline{G_{n}}%
(x,\omega )=$ $G_{n}^{\ast }(x,\omega )$ are biorthogonal (see Ref. \cite{MF} %
). The scalar product of modes $G_{n}(x,\omega )$ with $\overline{G_{m}}%
(x,\omega $)is simply given by:
\begin{equation}
\int_{0}^{\ell }G_{m}(x,\omega )G_{n}(x,\omega )dx=\Gamma
_{n}\delta _{nm} \label{65}
\end{equation}
where
\begin{equation}
\Gamma _{n}=\frac{\ell }{2}\left[ 1+c\frac{\sinh 2(i\theta
_{n}\ell /c+\varphi _{n})-\sinh 2\varphi _{n}}{2i\ell \theta
_{n}}\right] . \label{65a}
\end{equation}

Therefore modes $G_{n}(x,\omega )$ are orthogonal (for the product (\ref{65}%
)) and fulfill the same boundary conditions as $G(x,\omega )$,
contrary to
resonance modes $f_{n}(x)$ in Eq.~(\ref{3000}). Finally the solution of Eq.~(%
\ref{a3}) can be written as follows:
\begin{equation}
G(x,\omega )=c^{2}%
\mathrel{\mathop{\sum }\limits_{n}}%
\frac{G_{n}(x,\omega )G_{n}(x_{0},\omega )}{\Gamma _{n}(\theta
_{n}^{2}(\omega )-\omega ^{2})}.  \label{66}
\end{equation}

\subsubsection{ Calculation of poles and residues}

In order to calculate the inverse FT, the residue calculus will be
used again. The only terms of the series contributing to poles
verify:
\begin{equation}
\theta _{n}(\omega )=\pm \omega .  \label{66a}
\end{equation}

Looking at Eq.~(\ref{64}), it can be seen that these two solutions
lead to the same equation for $\omega $. Rewriting Eq.~(\ref{64})
by using Eqs.~(\ref {62}) and (\ref{63}), the resonance modes
frequencies are found to be solutions of Eq.~(\ref{20ab}).
Solutions $\omega _{p}$ of this equation are
the non zero poles of the integral in the inverse FT. Nevertheless the pole $%
\omega =0$ exists again, because the zero value satisfies
Eq.~(\ref{66a}), the eigenvalue $\theta _{n}(\omega )=0$
satisfying Eq.~(\ref{63a}).

It remains to calculate the residues. Starting with the poles
$\omega _{p}\neq $ $0$, we need to select in the series (\ref{66})
the terms involving poles. For a given $\omega _{p},$ there are
two terms. However it appears that modes corresponding to $\theta
_{n}$ and $-\theta _{n}$ are identical. As a consequence, only one
term of the series contributes to the inverse FT: it will be
denoted $\theta _{p}(\omega ).$ The corresponding residue is found
by expanding Eq.~(\ref{66}) for $\omega $ close to $\omega _{p}$,
as follows:

\begin{equation*}
G(x,\omega )=c^{2}\frac{G_{p}(x,\omega _{p})G_{p}(x_{0},\omega
_{p})}{\Gamma _{p}(2\omega _{p})(\omega _{p}-\omega )(1-\left[
\frac{d}{d\omega }\theta _{p}(\omega )\right] _{\omega =\omega
_{p}})}.
\end{equation*}

A similar expression can be found in Filippi \cite{filippi}, which
points out that Morse and Ingard \cite{morse} (p.559) forgot the
derivative. The same error is found in Morse and Feshbach
\cite{MF} (p.1347), with another error in the derivation of
Eq.~(\ref{63a}): these authors treated the problem of a string
with one non-rigid (and resistive) support.

Actually the derivative of $\theta _{p}(\omega )$, denoted $\theta
_{p}^{\prime }(\omega )$ can be calculated analytically, as
follows. Taking the derivative of Eq.~(\ref{64}) with respect to
$\omega ,$ or more conveniently, taking the logarithmic derivative
of Eqs.~(\ref{62}) and (\ref {63}), the following results are
obtained:

\begin{equation*}
\frac{1}{\omega _{p}}-\frac{\theta _{p}^{\prime }}{\theta _{p}}=\frac{%
2\varphi _{p}^{\prime }}{\sinh 2\varphi _{p}}= \frac{2(i\theta
_{p}^{\prime }\ell /c+\varphi _{p}^{\prime })}{\sinh 2(i\theta
_{p}\ell /c+\varphi _{p})}
\end{equation*}

Thus, eliminating the derivative $\varphi _{p}^{\prime }$, writing
$\theta _{p}=\omega _{p}$ and using Eq.~(\ref{65a}), it is found
after some algebra:
\begin{equation}
\;1-\theta _{p}^{\prime }=\ell/2\Gamma_{p}\,.  \label{81}
\end{equation}

Finally, for $\omega $ close to $\omega _{p},$:
\begin{equation}
G(x,\omega )=-\frac{c^{2}}{\ell }\frac{G_{p}(x,\omega
_{p})G_{p}(x_{0},\omega _{p})}{\omega _{p}(\omega -\omega _{p})},
\label{83}
\end{equation}

which is in accordance with Eq.~(\ref{24c}).

Otherwise, for $\omega $ close to $0$, the solution $\theta
(\omega )$ which is close to $0$, solution of Eq.~(\ref{66a}),
satisfies the following equation, deduced from (\ref{63a}) :
\begin{equation*}
\left[ 1-\Theta ^{2}/3+O(\Theta ^{4})\right] \left[ \Theta ^{2}+
\Omega ^{2}/(\zeta\zeta _{\ell })\right] =i\Omega \left[\zeta
^{-1}+\zeta _{\ell }^{-1} \right]
\end{equation*}

where $\Theta =\theta \ell /c$ and $\Omega =\omega \ell /c$. Therefore $%
\Theta ^{2}$ is of order $\Omega $, and
\begin{equation}
\theta ^{2}=i\omega c\ell ^{-1}\left[ \zeta ^{-1}+\zeta _{\ell
}^{-1}\right] +O(\omega ^{2}).  \label{70}
\end{equation}

Using Eq.~(\ref{66}), the residue for the pole $\omega =0$ is
obtained, and Eq.~(\ref{24d}) is confirmed. We conclude that the
method of the expansion in orthogonal modes in the Fourier domain
leads to the same result (Eq.~(\ref {29})) than the ``direct''
method, but the derivation is more delicate.

\subsection{System of two first-order equations}

We consider now the FT of (\ref{A20}):
\begin{equation}
i\omega {\bm \psi}(x,\omega )={\bm A}{\bm \psi}(x,\omega )+{\bm \phi }%
_{s}(x,\omega ),  \label{V1}
\end{equation}
where ${\bm \Psi}(x,\omega )=\text{ }(P(x,\omega ),V(x,\omega
))^{T}$. An interest of the system is that the boundary conditions
are independent of frequency:
\begin{equation}
P(0,\omega )=-\zeta V(0,\omega )\text{ ; }P(\ell ,\omega )=\zeta
_{\ell }V(\ell ,\omega ).
\end{equation}

Eigenvalues\ $\lambda _{n}$ and eigenvectors ${\bm \psi
}_{n}^{\alpha }$ of operator ${\bm A}$ are already known (see
Eqs.~(\ref{A5}) and (\ref{A6})). The appendix shows that the
adjoint operator of ${\bm A}$, is $\overline{{\bm A}}=-{\bm A}$,
and gives the boundary conditions for it. This formulation differs
slightly from the work \cite{TRO,TRO2}, considering a different
operator, but the principle is identical: we notice that these
authors treat the problem for more general operators and boundary
conditions. Eigenvalues and eigenvectors of $\overline{{\bm
A}\text{ }}$ are solutions of:
\begin{eqnarray*}
\overline{{\bm A}}\,\overline{{\bm \psi }_{m}^{\alpha }} &=&\overline{%
\lambda _{m}}\,\overline{{\bm \psi }_{m}^{\alpha }}\text{ \ \ } \\
\overline{p_{m}} &=&\zeta \overline{v_{m}}\text{ \ for }x=0\text{\; \ }%
\overline{p_{m}}=-\zeta _{\ell }\overline{v_{m}}\text{ \ for
}x=\ell .
\end{eqnarray*}

Thus, the adjoint eigenvalue problem to be solved is the same as
the direct
one, by replacing $c$ by $-c$, $\eta $ by -$\eta $ and $\eta _{\ell }$ by $%
-\eta _{\ell }$. The eigenelements are thus found to be:
\begin{equation}
\,\overline{{\bm \psi }_{m}^{\alpha }}=\left(
\begin{array}{c}
\overline{p_{m}}(x) \\ \overline{v_{m}}(x)
\end{array}
\right) =\left(
\begin{array}{c}
\cosh (\overline{\lambda _{m}}x/c+\eta ) \\ \sinh
(\overline{\lambda _{m}}x/c+\eta )
\end{array}
\right)
\end{equation}
where
\begin{equation}
\overline{\lambda _{m}}=-\left( \eta +\eta _{\ell }+im\pi \right)
c/\ell =\lambda _{-m}  \label{V3}
\end{equation}

Comparing with the family ${\bm\psi }_{n}^{\alpha }$ (Eq.
(\ref{A5})), there is a difference in sign for the second row: we
notice that Rideau\cite {rideau} made an error in the biorthogonal
family. By construction, the biorthogonality relationship is
ensured:
\begin{equation}
(\lambda _{n}-\overline{\lambda _{m}}^{\ast })<{\bm\psi }_{n}^{\alpha },%
\overline{{\bm\psi }_{m}^{\alpha }}>=0.
\end{equation}
Using Eq.~(\ref{J3}), we remark that $\lambda _{n}=\overline{\lambda _{m}}%
^{\ast }$ implies $m=-\nu ,$ as defined in Eq.~(\ref{J3nu}).
Therefore
\begin{equation}
<{\bm\psi }_{n}^{\alpha },\overline{{\bm\psi }_{m}^{\alpha }}%
>=\int_{0}^{\ell }\overline{{\bm\psi }_{m}^{\alpha }}^{T\ast }{\bm\psi }%
_{n}^{\alpha }\,dx=(-1)^{\mu }\,\ell \,\delta _{m,-\nu }.
\label{V4}
\end{equation}
 This latter relation enables to
perform a modal decomposition on the $\left( {\bm
\psi}_n^{\alpha}\right)$ family: but, contrarily to standard
cases, the $n$-th coefficient is not given by the scalar product
with ${\bm \psi}_n^{\alpha}$, but by the scalar product with
$\overline{{\bm \psi }_{-\nu_n}^{\alpha }}$, (up to the
normalization
coefficient $(-1)^{\mu}\,\ell$). Notice that if $\eta $ and $\eta _{\ell }$ are both real, $%
\overline{\lambda _{m}}=\lambda _{m}^{\ast }$, and (\ref{V4}) is
obvious
from the expressions of eigenvalues and eigenvectors (for this case, $%
\lambda _{n}=\overline{\lambda _{m}}^{\ast }$ implies $n=m).$ For
the
general case, the scalar product can be written: $<{\bm%
\psi }_{n}^{\alpha },\overline{{\bm\psi }_{m}^{\alpha
}}>=\int_{0}^{\ell }\cos \left[ \beta _{n}(x)+\beta
_{-m}(x)\right] dx=(-1)^{\mu }\int_{0}^{\ell }\cos \left[ ({\nu
}+m)\pi x/\ell \right]dx$. Comparison with Eq.~(\ref{A17})
exhibits the difference between the two methods. \\
 It remains to
apply orthogonality to Eq.~(\ref{V1}). We choose the case of the
Green function (Eqs.~(\ref{A221})), with the following result:
\begin{equation}
G(x,\omega )=-\frac{c^{2}}{\ell }%
\mathrel{\mathop{\sum }\limits_{n}}%
\frac{f_{n}(x)f_{x}(x_{0})}{\omega (\omega -\omega _{n})}.
\end{equation}
The calculation is easy, because $\omega _{n}$ does not depend on
frequency. Comparison with Eq.~(\ref{3000}) exhibits a difference
in the denominator, i.e. a factor $\omega $ instead of $\omega
_{n}$, and, of course, the absence of constant mode. When
returning to the time domain, all the terms corresponding to
$\omega =\omega _{n}$ are identical, and a constant mode is found
for the pole $\omega =0$, but again it is not possible to deduce
it from orthogonality relations, as in section \ref{ortho}.
Nevertheless, because of the independence of the boundary
conditions with respect to frequency, the calculation of the
residues is much easier than for the second-order equation. For
the same reason, the calculation in the time domain would be
possible with the same modal decomposition, and this is a major
difference with the methods based upon the second-order equation.

\section{Conclusion}

The simple problem we have studied, which can be regarded in
particular as a radiation problem, exhibits interesting properties
for the resonance modes:
they are complex-valued, and non orthogonal for the simple product (\ref{35a}%
) because of the bounded character of the considered medium, but
except the constant mode, they are orthogonal for a product
modified in a proper way, and are a basis for the space of
solutions. Second order equations allow to find the constant mode,
while first order systems of equations allow a more direct
formulation of boundary impedances.

Thanks to the simplicity of the problem, the analytical treatment
is possible with several methods, elucidating the relationship
between them, which can be useful for more intricate problems
(e.g. when damping is added to propagation, or when boundary
impedances involve a mass or a spring). No approximations are
needed, the results are valid whatever the value of the terminal
resistances. Active boundaries can also be considered, thanks to a
change in functions. We notice that an advantage of the frequency
domain calculations is the possibility of the treatment of an
arbitrary dependence of the boundary conditions. For a dependence
$\eta (\omega )$ and $\eta
_{\ell }(\omega )$, Eq. (\ref{3000}) remains valid by replacing $\ell $ by $%
\left[\ell -ic(\eta _{p}^{\prime }+\eta _{\ell p}^{\prime
})\right]$, where $\eta
_{p}^{\prime }=(d\eta /d\omega )_{\omega =\omega _{p}},$ and similarly for $%
\eta _{\ell }.$ This can be shown by generalyzing Eq. (\ref{23b}),
or, with some algebra, using the modal expansion.

Finally, considering the problem of a stratified medium (see
section \ref
{statement}), it could be deduced in the field outside of the interval $%
\left[ 0,\ell \right] .$ When terminations are passive, a result
is that modes tend to infinity when $x$ tends to $\pm \infty $. An
interesting study has been done in Ref. \onlinecite{leung}, using
biorthogonality and explaining the relation between the energy
outside the interval and the terms responsible of non
orthogonality in equation (\ref{433}).
\section*{Acknowledgements}

We would like to thank Jos\'{e} Antunes, Patrick Ballard, Sergio
Bellizzi, Michel Bruneau, Paul Filippi, Dominique Habault,
Pierre-Olivier Mattei and Vincent Pagneux for very fruitful
discussions.

\appendix

\section{Proof of the completeness of the eigenelements of operator   ${\bm A}$}

The operator${\bm A}$, defined by Eq.(\ref{A3}) is a differential operator defined on the energy space $H=L^{2}(0,\ell )\times L^{2}(0,\ell )$, it has a compact resolvent (cf Ref. %
\onlinecite{int}, p.191). Using the ordinary scalar product $<{\bm
\psi }, {\bm \varphi >=}\int_{0}^{\ell }\left[ pq^{\ast }+vw^{\ast
}\right] dx$, between ${\bm \psi }(x,t)=(p,v)^{T}$ and ${\bm
\varphi }(x,t)=(q,w)^{T}$, the following result is obtained:
\begin{eqnarray*}
&<&{\bm A}{\bm \psi },{\bm \varphi }>+<{\bm \psi },{\bm A}{\bm
\varphi >=} \\
&&{\bm -}c\left[ v(q^{\ast }+\zeta _{\ell }w^{\ast })\right] _{x=\ell }+c%
\left[ v(q^{\ast }-\zeta w^{\ast })\right] _{x=0}.
\end{eqnarray*}

It is deduced that the adjoint operator of ${\bm A}$ is $\overline{{\bm A}}=-%
{\bm A}$ (we denote all quantities related to the adjoint problem
with an overline), and on its domain, the following adjoint
boundary conditions must be fulfilled:
\[
q(0,t)=\zeta ^{\ast }w(0,t)\text{ \ and \ }q(\ell ,t)=-\zeta
_{\ell }^{\ast }w(\ell ,t)\text{ \ }\forall t
\]
(here $\zeta =\zeta ^{\ast },$ and $\zeta _{\ell }=\zeta _{\ell
}^{\ast }$;
if $\zeta $ is infinite, the boundary conditions are $v(0,t)=0$, and $%
w(0,t)=0,$ and similarly for boundary $x=\ell $). Therefore ${\bm
A}$ is
skew-{\em symmetric}, but not skew-{\em adjoint}, because the domains of ${\bm A}$ and $%
\overline{{\bm A}}$ are different, except if both $\zeta $ or
$\zeta _{\ell } $ are either zero or infinite (Dirichlet or
Neumann conditions); notice that for a skew-adjoint operator, the
eigenvalues are purely imaginary. In order to find a new scalar
product, we denote, from Eqs. (\ref{A7}) to (\ref{A9}):

\[
{\bm \psi }_{n}^{\alpha }(x)=\left(
\begin{array}{cc}
e^{\alpha (x)} & e^{-\alpha (x)} \\ -e^{\alpha (x)} & e^{-\alpha
(x)}
\end{array}
\right) \left(
\begin{array}{c}
e^{i\beta _{n}(x)} \\ e^{-i\beta _{n}(x)}
\end{array}
\right) .
\]

In $H$ the standard scalar product $<%
{\bm \psi }_{n}^{\alpha },{\bm \psi }_{p}^{\alpha }>_{H}=\int_{0}^{\ell }%
\left[ p_{n}p_{p}^{\ast }+v_{n}v_{p}^{\ast }\right] dx$ does not vanish for $%
n\neq p$, except if $\ \alpha (x)=0.$ If we denote ${\bm \psi
}_{n}^{0}(x)$ the functions corresponding to the latter case, it
is possible to construct a new scalar product ensuring
orthogonality, in a similar way Rideau \cite{rideau} did. From Eq.
(\ref{A12}), the following {\it hyperbolic rotation} is obtained:
\begin{equation}
{\bm \psi }_{n}^{0}(x)={\bm G}_{\alpha }(x){\bm \psi }_{n}^{\alpha
}(x).
\end{equation}
 We will now prove that the new product
\begin{equation}
<{\bm \psi },{\bm \varphi }>_{H}^{\alpha }=<{\bm G}_{\alpha }{\bm \psi },%
{\bm G}_{\alpha }{\bm \varphi }>_{H}=\int_{0}^{\ell }{\bm \varphi}^{T\ast }\,%
{\bm M}_{\alpha}(x)\,{\bm \psi}\,dx
\end{equation}
where ${\bm M}_{\alpha }(x)={\bm G}_{\alpha }^{T}{\bm G}_{\alpha
}$, leads to the orthogonality of the modes. ${\bm M}_{\alpha
}(x)$ is found to be equal to ${\bm G}_{2\alpha }$. It is
symmetrical and positive definite because
\begin{eqnarray*}
(\left\| {\bm \psi }\right\| _{H}^{\alpha })^{2} =(\left\|
(p,v)^{T}\right\| _{H}^{\alpha })^{2}= \\ \int_{0}^{\ell }\left[
\cosh \left[ 2\alpha (x)\right] (\left| p\right|
^{2}+\left| v\right| ^{2})+2\sinh \left[ 2\alpha (x)\right] \Re e(pv^{\ast })%
\right] dx
\end{eqnarray*}

can be rewritten as:
\[
(\left\| {\bm \psi}\right\| _{H}^{\alpha })^{2}=\frac{1}{2}\int_{0}^{\ell }%
\left[ e^{2\alpha (x)}\left| p+v\right| ^{2}+e^{-2\alpha
(x)}\left| p-v\right| ^{2}\right] dx.
\]

Moreover $\alpha (x)$ is a function varying monotonously from
$\eta _{r}$ to $-\eta _{\ell r}$ when $x$ increases from $0$ to
$\ell $, and the following bounds can be found for $\left\| {\bm
\psi}\right\| _{H}^{\alpha }$:
\begin{equation}
c_{\alpha }\left\| {\bm \psi}\right\| _{H}<\left\| {\bm
\psi}\right\| _{H}^{\alpha }<C_{\alpha }\left\| {\bm \psi}\right\|
_{H}  \label{A15}
\end{equation}
where $c_{\alpha }=e^{-\widetilde{\eta }}$ and $C_{\alpha }=e^{\widetilde{%
\eta }}$, with $\widetilde{\eta }=\sup \left[ \left| \eta
_{r}\right| ,\left| \eta _{\ell r}\right| \right] $.

Therefore the modes ${\bm \psi }_{n}^{\alpha }$ are orthogonal for
the new scalar product $<{\bm \psi },{\bm \varphi }>_{H}^{\alpha
}.$ First recall that $\left( {\bm \psi }_{n}^{0}\right) _{n}$ is
the family of eigenvectors of a classically skew-adjoint operator
with compact resolvent, it is thus complete in $H$. Now, thanks to
(\ref{A15}), the two norms are equivalent on
$H$, and the hyperbolic rotation shows that $\left( {\bm \psi }%
_{n}^{0}\right) _{n}$ and $\left( {\bm \psi }_{n}^{\alpha }\right)
_{n}$ span the same subspace, namely the whole of $H$. This proves
the completeness of $\left( {\bm \psi }_{n}^{\alpha }\right) _{n}$
in $H$.

The calculation leads to the simple result:
\begin{eqnarray}
&<&{\bm \psi }_{n}^{\alpha },{\bm \psi }_{m}^{\alpha }>_{H}^{\alpha }=<{\bm %
\psi }_{n}^{0},{\bm \psi }_{m}^{0}>_{H}=\int_{0}^{\ell }\cos
\left[ \beta _{n}(x)-\beta _{m}(x)\right] dx  \nonumber \\
&=&\int_{0}^{\ell }\cos \left[ (n-m)\pi x/\ell \right] dx=\ell
\delta _{nm}. \label{A17}
\end{eqnarray}

\bibliography{apssamp-asa}
\bibliographystyle{jasasty}

\end{document}